\documentclass{aa5}
\usepackage{epsf}
\newcommand{\etal}{{et al}\/.}
\newcommand{\new}[1]{{#1}}
\begin{document}
\headnote{Research Note}
\title{An optical inverse-Compton hotspot in 3C\,196?}
\author{M.J. Hardcastle}
\institute{Department of Physics, University of Bristol, Tyndall
Avenue, Bristol BS8 1TL, UK (m.hardcastle@bristol.ac.uk)}
\date{Version of \today}

\abstract{Several hotspots of FRII radio sources have previously been
detected in the X-ray at a flux level consistent with the X-rays being
due to inverse-Compton scattering of radio synchrotron photons
(`synchrotron self-Compton'), if the magnetic fields in the hotspots
are close to their equipartition values. However, the number of
hotspots compact and bright enough to exhibit detectable X-ray
emission is small, so it is worth searching for synchrotron
self-Compton emission in the optical, in spite of the obvious
observational difficulties of such an approach. In this note I report
on a possible detection of an optical inverse-Compton hotspot in a
deep {\it Hubble Space Telescope} observation of the distant quasar
3C\,196, at a level which implies a hotspot magnetic field strength
close to equipartition if the electrons have a low-energy cutoff
around $\gamma \sim 500$.  \keywords{galaxies: active -- radiation
mechanisms: non-thermal -- quasars: individual: 3C\,196} } \maketitle

\section{Introduction}

The relativistic electron population responsible for synchrotron
emission in extragalactic radio sources necessarily scatters incoming
photons up to higher energies by the inverse-Compton process. Possible
source photon populations include the microwave background, starlight
from the host galaxy, photons from the active nucleus and the
synchrotron photons themselves; different populations will dominate in
different regions of the source. In the compact hotspots of powerful
double (FRII) radio sources, the dominant component is expected to
be the synchrotron photons, and the so-called `synchrotron
self-Compton' (SSC) process should accordingly dominate the inverse-Compton
emissivity.

Observations of SSC emission are important because they allow us to
make a measurement of the magnetic field strength in the
hotspot. We know the synchrotron flux density and the dimensions
of the hotspot, which give us the synchrotron emissivity and photon
energy density. A measurement of the inverse-Compton emissivity then
tells us about the electron energy density, allowing us to infer the
magnetic field strength from the observed synchrotron emission.

However, SSC emission from hotspots is expected to be faint. So far
there are four convincing cases of inverse-Compton emission detected
with X-ray observations, in the radio galaxies 3C\,405 (Harris,
Carilli \& Perley 1994; Wilson, Young \& Shopbell 2000), 3C\,295
(Harris \etal\ 2000), 3C\,123 (Hardcastle \etal\ 2000) and the quasar
3C\,263 (Hardcastle \etal\ in preparation); these detections are
consistent with a hotspot magnetic field strength close to the minimum
energy or equipartition value. These objects represent
some of the brightest well-studied hotspots in the sky, and the
faintest of them approaches the detection limit of long {\it Chandra}
observations. There may be fewer than ten objects in the entire sky
that have hotspots whose SSC emission is detectable at a useful level
in the X-ray with the present generation of instruments.

Because the spectrum of SSC emission is expected to be similar to that
of synchrotron emission, SSC emission in the optical ought to be
detectable at flux levels higher than those seen in the
X-ray. However, there are several difficulties with observing this in
practice. Firstly, in many cases, the high-frequency radio/sub-mm/IR
spectrum of hotspots is poorly known, which means that it is hard to
say whether an optically detected hotspot is synchrotron or
inverse-Compton --- a number of sources have well-known optical
synchrotron hotspots. For example, it is not clear whether the optical
hotspot of 3C\,295 detected by Harris \etal\ (2000) is synchrotron or
SSC in nature. Secondly, because the increase in frequency in the SSC
process goes as $\gamma^2$, SSC at low frequencies probes low-energy
electrons (with $\gamma \la 1000$) and we typically do not know much
about the radio emission from these electrons; our models are
accordingly uncertain. And finally there are practical difficulties;
emission from the hotspots is often too faint to be seen in the
optical against the background from the host galaxy or active nucleus.
However, detection of optical SSC emission is still possible in
principle, and could give us valuable information about the magnetic
field strengths and low-energy electron populations in hotspots. In
this note I report on a possible detection of optical SSC
emission from the distant quasar 3C\,196. Except where otherwise
stated, I use a cosmology with $H_0 = 65$ km s$^{-1}$ Mpc$^{-1}$,
$\Omega_{\rm m} = 0.3$, $\Omega_\Lambda = 0.7$.

\section{Observations}

\begin{figure*}
\epsfxsize 15cm
\begin{center}
\epsfbox{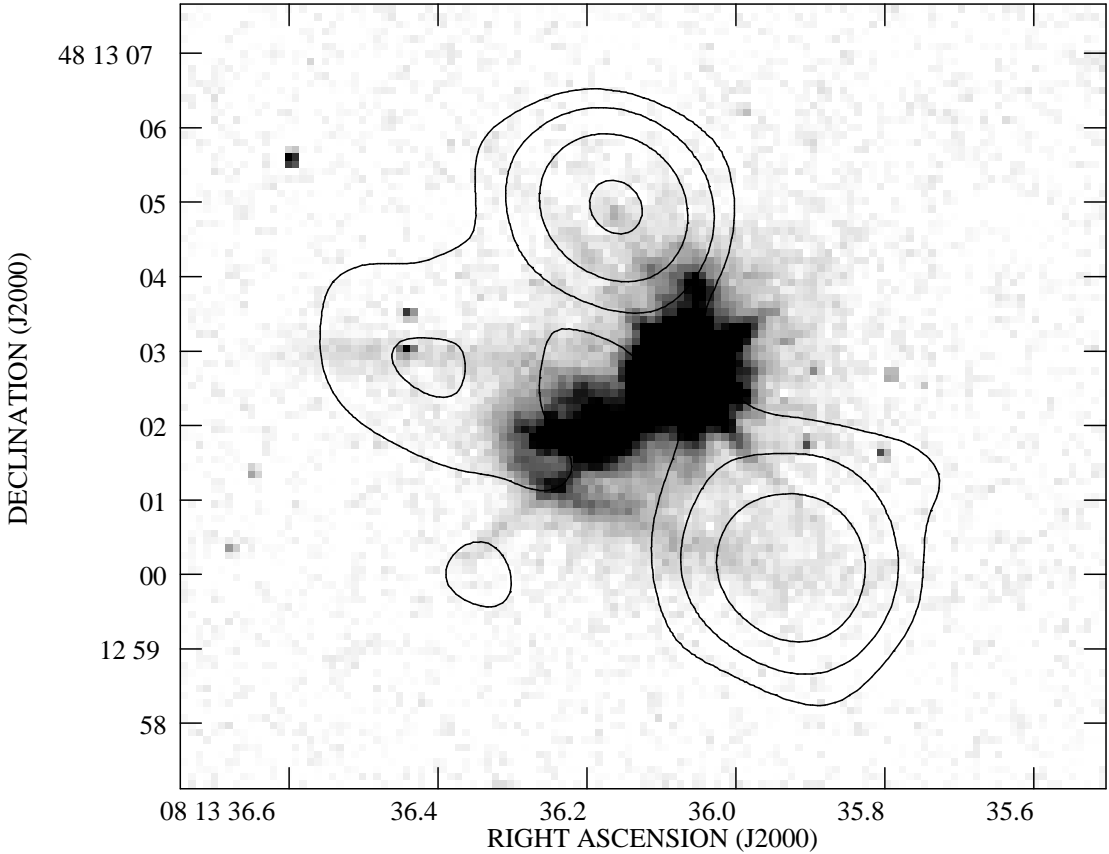}
\end{center}
\caption{The WFPC2 (WFC3) image of 3C\,196. The linear greyscale range
is from 144 to 250 counts per pixel. The PSF has not been subtracted,
and its diagonal diffraction spikes can be seen. Below the quasar to
the SE is the foreground spiral galaxy. The apparent `box' to the W of
the quasar is the result of a bad pixel on the CCD, and illustrates
the dithering strategy used by RS. \new{Radio contours are at $5 \times (1,
4, 16, 64)$ mJy beam$^{-1}$, taken from a 15-GHz VLA map with a
resolution of $1.34 \times 1.09$ arcsec.} The {\it HST} positions have
been corrected to align the quasar with the radio core.}
\label{image}
\end{figure*}

\begin{figure}
\epsfxsize 8.8cm
\epsfbox{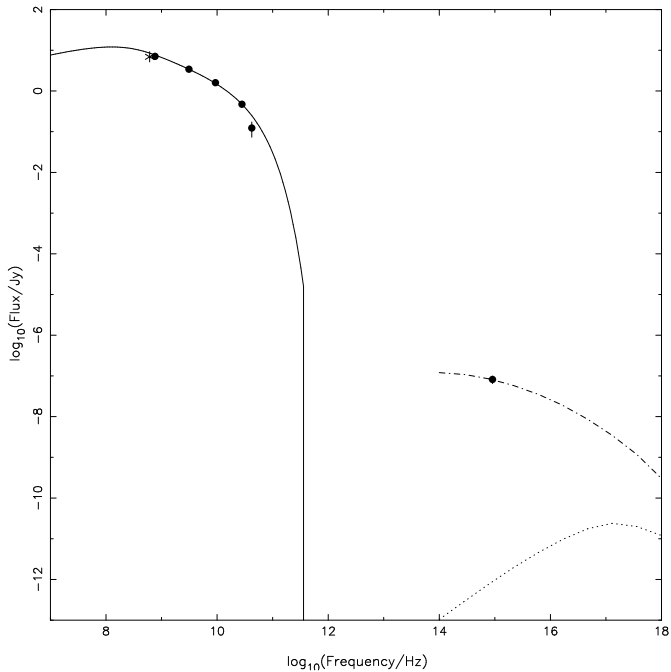}
\caption{The synchrotron and inverse-Compton spectrum of 3C\,196's N
hotspot. Radio data are from Table \ref{fluxes}, and are plotted in
the rest frame of the quasar. The optical data
point is discussed in the text. The dot-dashed line shows
a synchrotron self-Compton spectrum fitted to the data, using a model with
equipartition magnetic field strength and $\gamma_{\rm min} =
420$. The dotted line is the contribution from inverse-Compton
scattering of the microwave background for these parameters.}
\label{flux}
\end{figure}

3C\,196 is a $z=0.871$ quasar with a compact, FRII-like radio
structure (Laing 1982; Lonsdale \& Morison 1983, hereafter LM; Brown
1990). It came to our attention as a possible target for {\it Chandra}
observations because of its very bright, compact hotspots, with flux
densities around 2 Jy at 5 GHz (LM). However, investigation of the
available high-frequency data in the VLA archive\footnote{The National
Radio Astronomy Observatory Very Large Array (VLA) is operated by
Associated Universities Inc., under co-operative agreement with the
National Science Foundation.} showed that the radio spectrum of the
hotspots cuts off very steeply at observing frequencies of tens of GHz
(Table \ref{fluxes}, Fig.\ \ref{flux}). Because it is photons at
around these frequencies that are scattered up to X-ray energies,
3C\,196 became less attractive as a {\it Chandra} target. But the
cutoff in the synchrotron spectrum does rule out the possibility of an
optical synchrotron hotspot. For this reason it is interesting to look
for optical inverse-Compton emission in this object.

\begin{table}
\caption{Radio flux densities for the hotspots of 3C\,196}
\label{fluxes}
\begin{center}
\begin{tabular}{lrrl}
\hline
Frequency&North&South&Reference\\
(GHz)&(Jy)&(Jy)\\
\hline
0.329&6.9&--&1\\
0.408&$7.1\pm 0.4$&$15.2\pm0.6$&2\\
1.67&$3.420\pm 0.2$&$5.58\pm 0.3$&2\\
5.0&$1.6\pm0.2$&$2.5 \pm 0.2$&2\\
15.0&$0.472\pm 0.02$&$0.47\pm 0.07$&3\\
22.5&$0.123\pm 0.05$&--&3\\
\hline
\end{tabular}
\vskip 8pt
\begin{minipage}{7.6cm}
\new{Data tabulated are for the compact component of the hotspot,
which in most cases was unresolved.} References are: (1) Linfield \&
Simon (1984) (2) LM (3) This paper, from VLA archive. The N hotspot is
partially resolved at 22 GHz, so the flux quoted may be an
underestimate.
\end{minipage}
\end{center}
\end{table}

3C\,196 has already been well studied in the optical (e.g.\ Boiss\'e
\& Boulade 1990; Cohen \etal\ 1996), because the quasar lies close to
a $z=0.437$ barred spiral galaxy which gives rise to absorption in HI
(Brown \& Mitchell 1983; Brown \etal\ 1988; \new{Briggs, de Bruyn \&
Vermeulen 2001}) and optical lines (Foltz,
Chaffee \& Wolfe 1988). Because the hotspots lie only $\sim 2$ arcsec
from the quasar nucleus, high resolution is needed to separate any possible
optical hotspot emission from the nucleus. The deepest {\it Hubble
Space Telescope} ({\it HST}) image is presented by Ridgway \& Stockton (1997;
hereafter RS), and consists of 8 dithered observations of 900 s
duration each on the WFC3 chip in the F622W filter. After combining
the images and subtracting the PSF, they find some extended emission
probably related to the quasar, as well as imaging the foreground
spiral galaxy, but do not comment on any possible hotspot emission.

I have obtained the data of RS from the {\it HST} archive and combined the
individual observations, using the IRAF task {\sc crrej} to remove
cosmic rays in the pairs of observations with the same pointings
followed by the AIPS tasks {\sc hgeom} and {\sc comb} to stack the
four dither directions. In Fig. \ref{image} I show a greyscale of the
resulting image. There is a weak but clear source coincident with the
northern hotspot, a component which can also be seen on the image of
RS. One of the spiral arms of the foreground galaxy crosses the
position of the southern hotspot, rendering any discussion of that
component impossible.

The flux density of the component coincident with the northern hotspot
can be determined by small-aperture photometry. After subtraction of
the well-determined sky background and of a locally determined (and
more uncertain) correction for the background local to the source, I
find the source to contain $133\pm 27$ counts in an extraction region
with a radius of 3 pixels. Correcting for the effects of the PSF
(Holtzman \etal\ 1995) and for a small amount of Galactic reddening
[$E(B-V) = 0.058$, according to the data of Schlegel, Finkbeiner \&
Davis (1998)] this translates, using factors provided by the IRAF {\sc
synphot} package, to a flux density of $82 \pm 17$ nJy at an observing
frequency of $4.85 \times 10^{14}$ Hz.

It is worth briefly considering whether additional reddening might be
introduced by the foreground spiral galaxy. The northern hotspot is
the component of 3C\,196 furthest from the spiral; the separation of
3.1 arcsec corresponds to a distance of 23 kpc at the redshift of the
spiral. Observations of optical line absorption against
the quasar nucleus show that absorbing material from the spiral
certainly has an effect at $\sim 10$ kpc. On the other hand, we know
from the VLBI work of Brown \etal\ (1988) that the HI column density
to the hotspot is $\la 3 \times 10^{20}$ cm$^{-2}$, which would imply
$E(B-V)$ in the frame of the galaxy of $\la 0.06$; this could give up
to $0.14$ mag of reddening at our observing wavelength, but this would
only increase the inferred flux density by 14 per cent, to 93 nJy. The
effect is therefore not significant. \new{The detailed models of Briggs
\etal\ (2001) place the northern hotspot outside the absorbed region.}

\section{Inverse-Compton?}

In order to calculate the inverse-Compton flux density expected at
this frequency we must assume a size and geometry for the hotspot. To
carry out the calculation I use the code of Hardcastle, Birkinshaw \&
Worrall (1998) which assumes that the hotspot is a homogeneous
sphere. The radius of the sphere is set to 0.3 arcsec, based on the
MERLIN image of Lonsdale (1984).  The synchrotron spectrum is then fit
with a simple model consisting of a low-frequency power law with
spectral index 0.5 and a high-energy cutoff; this constrains the upper
energy of the electrons, and gives an adequate fit to the radio data
(Fig.\ \ref{flux}). The unknown parameters are then the low-energy
cutoff of the electron spectrum, and the magnetic field strength in
the hotspot.

\begin{figure}
\epsfxsize 8.8cm
\epsfbox{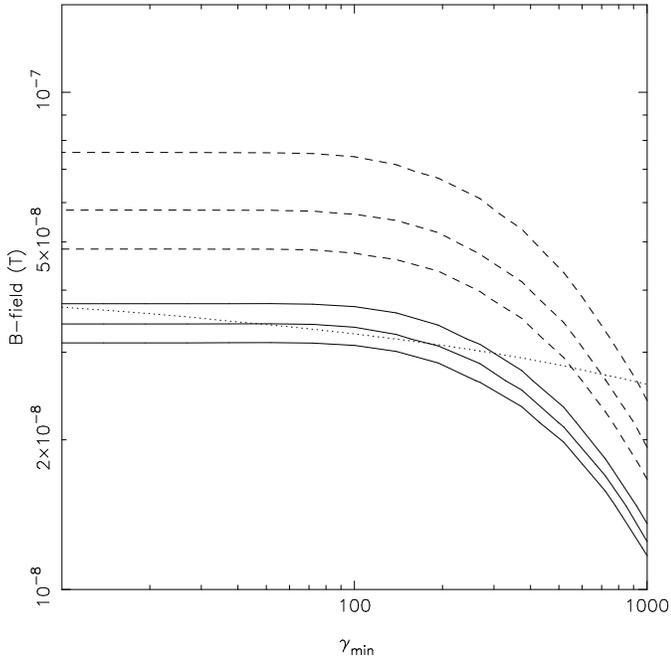}
\caption{The difference between predicted and observed values of the
optical flux as a function of $\gamma_{\rm min}$ and $B$. The contours
show a difference between prediction and observation of 1, 2 and 3
$\sigma$; \new{the solid contours show the model {\it overpredicting}
the observed optical flux, while the dashed contours show where the
model {\it underpredicts} the observations.} The dotted line shows the
equipartition field strength as a function of $\gamma_{\rm min}$. The
values plotted here are for the default cosmology. Other standard
choices of cosmological parameters (e.g. $H_0 = 50$ km s$^{-1}$
Mpc$^{-1}$, $\Omega_{\rm m} = 0.0$, $\Omega_\Lambda = 0.0$) give very
similar results.}
\label{sigmacont}
\end{figure}

\new{The value of the low-energy cutoff has a strong effect on the
predicted optical SSC emissivity, because it is the low-energy
electrons that scatter radio photons into the optical band. For a
given energy density in photons, high cutoffs reduce the optical
emissivity, while low cutoffs increase it. A smaller but still
significant effect is that a lower cutoff implies a greater energy
density in electrons, giving rise to an increased equiapartition
magnetic field strength. The fact that the source
is detected at 408 MHz and below constrains the low-energy cutoff;
$\gamma_{\rm min} \la 800 (B_{\rm eq}/B)^{1\over2}$, where $B_{\rm
eq}$ is the equipartition magnetic field strength assuming no protons
and filling factor unity. The best fits to the entire radio spectrum
are given by $\gamma_{\rm min} \approx 600$, which is comparable to
the low-energy cutoffs inferred in Cygnus A and 3C\,123 (Table
\ref{compare}).

To see what regions of $B$ and $\gamma_{\rm min}$ are consistent with
the observed optical emission, I allowed them both to vary over a wide
range and determined the difference between the predicted and observed
optical flux density. The results are plotted in Fig.\ \ref{sigmacont}.}

It will be seen that even if we treat the optical emission from the
hotspot as unrelated to the inverse-Compton process, and use it to
give an upper limit, this calculation constrains the magnetic field
strength in the hotspot. The region in the bottom left of Fig.\
\ref{sigmacont}, \new{below the solid contours}, is excluded at better
than the $3\sigma$ level by the data, \new{since parameters in this
region would produce more optical emission than is observed}. This
implies, for plausible $\gamma_{\rm min}$, that the magnetic field
strength \new{in the hotspot} is close to or greater than the
equipartition value. If we believe that inverse-Compton emission has
actually been detected in this object, then the magnetic field
strength implied is very close to the equipartition value if
$\gamma_{\rm min} \sim 400$ -- $600$, a plausible fit to the data. If
the low-energy cutoff in the electron spectrum is much lower than
this, then to avoid producing an optical SSC hotspot brighter than
that observed we require magnetic field strengths greater than the
equipartition value, \new{although the difference is never very large,
and approaches $1\sigma$ for $\gamma_{\rm min} \approx 10$}; but the
radio data are less well fitted by \new{models with low $\gamma_{\rm
min}$. Table \ref{compare} compares the
derived field strengths for 3C\,196, on the assumption of $\gamma_{\rm
min} = 600$, with those obtained for other
sources. The results are broadly similar.}

\begin{table}
\new{
\caption{Inverse-Compton parameters obtained for 3C\,196 and other
sources.}
\label{compare}
\begin{tabular}{lrrrrl}
\hline
Source&$\gamma_{\rm min}$&$u_\nu$ ($\times 10^{-11}$&$B_{\rm eq}$&$B_{\rm SSC}$&Ref.\\
&&J m$^{-3}$)&(nT)&(nT)\\
\hline
3C\,405 A&420&1.0&31&27&1\\
3C\,295 N&800&8.2&63&30&2\\
3C\,123&1000&0.57&17&12&3\\
3C\,196&600&1.3&28&24&4\\
\hline
\end{tabular}
\vskip 5pt
\begin{minipage}{\linewidth}
I tabulate information for hotspot A of Cygnus A and the N hotspot of
3C\,295. $\gamma_{\rm min}$ is the value assumed in the calculation;
see the text for a discussion of the values applicable to 3C\,196.
$u_\nu$ is the mean energy density in synchrotron photons; $B{\rm eq}$
is the equipartition magnetic field strength, using assumptions given
in the text; $B_{\rm SSC}$ is the field required if the X-ray or
optical emission observed is all to be modelled as SSC. References
are: (1) Harris \etal\ (1994); (2) Harris \etal\ (2000); (3)
Hardcastle \etal\ (2001); (4) this paper. For consistency,
all calculations have been carried out using the Hardcastle \etal\
(1998) code and the cosmological parameters of this paper, so the
results quoted here differ slightly from published values.
\end{minipage}
}
\end{table}

As Brunetti, Setti \& Comastri (1997) have pointed out, photons from
the active nucleus can also be inverse-Compton scattered to higher
energies by electrons in the radio components. The N hotspot in
3C\,196 is a projected distance of only 20 kpc from the nucleus, and
so the energy density of nuclear photons in the hotspot may be
significant. However, the restricted range of electron energies
inferred in the hotspot ($500 \la \gamma \la 6000$) mean that only
photons in the radio band can be scattered into the optical. The radio
nucleus of 3C\,196, as we observe it, is unusually weak for a quasar
--- only 7 mJy at 5 GHz (Reid \etal\ 1995). Taking into account the
effects of beaming in the manner discussed by Hardcastle \etal\ (2001)
I find that, if the quasar is at an angle of less than 45 degrees to
the line of sight, extremely high nuclear bulk Lorentz factors ($\ga
25$) are required to make the number spectral density of nuclear radio
photons equal to that seen in the hotspot. A model of this sort seems
unlikely to be viable for this particular source.

\section{Conclusions}

A weak optical component coincident with the northern radio hotspot is
detected in 3C\,196. The radio spectrum makes this component very
unlikely to be due to synchrotron emission. If it is used as
an upper limit on any optical SSC emission, it requires the magnetic
field strength in the hotspot to be greater than or equal to the
equipartition value. If it is taken to be a {\it detection} of SSC emission,
its flux level is in good agreement with a model similar to the one
found to work in X-ray detected hotspots; the low-energy cutoff is
around $\gamma = 500$ and the magnetic field strength is close to the
equipartition value.

This work illustrates the possibility of finding optical
inverse-Compton hotspots in deep observations of radio sources.
At the time of writing I am not aware of any other sources with bright
compact hotspots of which suitable optical observations exist;
observers are encouraged to be alert to the possibility of finding
such components in their data.

\begin{acknowledgements}
I thank Dan Harris for a helpful referee's report on the first version
of this paper.
\end{acknowledgements}


\begin{thebibliography}{}
\bibitem[]{99}Boiss\'e P., Boulade O., 1990, A\&A, 236, 291

\bibitem[]{121}Briggs F.H., de~Bruyn A.G., Vermeulen R.C., 2001, A\&A in~press astro-ph/0104457

\bibitem[]{124}Brown R.L., 1990, in Zensus J.A., Pearson T.J., eds, Parsec-scale Radio Jets, Cambridge University Press, Cambridge, p.~199

\bibitem[]{125}Brown R.L., Mitchell K.J., 1983, ApJ, 264, 87

\bibitem[]{126}Brown R.L., Broderick J.J., Johnston K.J., Benson J.M., Mitchell K.J., Waltman E.B., 1988, ApJ, 329, 138

\bibitem[]{131}Brunetti G., Setti G., Comastri A., 1997, A\&A, 325, 898

\bibitem[]{165}Carilli C.L., Perley R.A., Dreher J.W., Leahy J.P., 1991, ApJ, 383, 554

\bibitem[]{183}Cohen R.D., Beaver E.A., Diplas A., Junkkarinen V.T., Barlow T.A., Lyons R.W., 1996, ApJ, 456, 132

\bibitem[]{290}Foltz C.B., Chaffee F.H., Wolfe A.M., 1988, ApJ, 335, 35

\bibitem[]{342}Hardcastle M.J., Birkinshaw M., Worrall D.M., 1998, MNRAS, 294, 615

\bibitem[]{343}Hardcastle M.J., Birkinshaw M., Worrall D.M., 2001, MNRAS, 323, L17

\bibitem[]{355}Harris D.E., Carilli C.L., Perley R.A., 1994, Nat, 367, 713

\bibitem[]{361}Harris D.E., et al., 2000, ApJ, 530, L81

\bibitem[]{383}Holtzman J., et~al., 1995, PASP, 107, 156

\bibitem[]{465}Laing R.A., 1982, in Heeschen, D.S., Wade C.M., eds, Extragalactic Radio Sources, IAU Symposium 97, Reidel, Dordrecht, p.~161

\bibitem[]{515}Linfield R., Simon R.S., 1984, AJ, 89, 1799

\bibitem[]{531}Lonsdale C.J., 1984, MNRAS, 208, 545

\bibitem[]{536}Lonsdale C.J., Morison I., 1983, MNRAS, 203, 833 [LM]

\bibitem[]{683}Ridgway S.E., Stockton A., 1997, AJ, 114, 511

\bibitem[]{697}Reid A., Shone D.L., Akujor C.E., Browne I.W.A., Murphy D.W., Pedelty J., Rudnick L., Walsh D., 1995, A\&AS, 110, 213

\bibitem[]{756}Schlegel D.J., Finkbeiner D.P., Davis M., 1998, ApJ, 500, 525

\bibitem[]{873}Wilson A.S., Young A.J., Shopbell P.L., 2000, ApJ, 544, L27

\end{thebibliography}
\end{document}